\documentclass{aa}
\usepackage{graphicx}
\usepackage{latexsym}
\def\rxj{RX\,J\,2117.1$+$3412}
\def\ngc{NGC\,246}
\def\k1{\mbox{K\,1-16}}
\def\kpd{KPD\,0005$+$5106}

\newcommand{\iion}[2]{\mbox{#1\ {\scriptsize #2}}}
\def\logg{$\log g$\thinspace}

\newcommand{\Teff}{\hbox{$T_{\rm eff}$}}

\def\etal{{et\thinspace al.}\ }

\def\kms{km\thinspace s$^{-1}$}

\def\deg{\ifmmode ^\circ \else $^\circ$ \fi }
\def\solar{\ifmmode _{\mathord\odot}\else $_{\mathord\odot}$\fi}
\begin{document}
   \title
   {Iron abundance in hot hydrogen-deficient central stars and white dwarfs
from FUSE, HST, and IUE spectroscopy\thanks{Based on observations obtained 
with the Far Ultraviolet Spectroscopic Explorer (FUSE), the Hubble Space 
Telescope (HST), and the International Ultraviolet Explorer (IUE)}}

   \titlerunning{Iron in hot H-deficient central stars and white dwarfs}

   \author{S. Miksa\inst{1,2}, J.L. Deetjen\inst{1}, S. Dreizler\inst{1}, J.W. Kruk\inst{3}, T. Rauch\inst{1}, \and K. Werner\inst{1}}
   \offprints{K. Werner}

   \authorrunning{S. Miksa \etal}
 
   \institute
    {Institut f\"ur Astronomie und Astrophysik, Universit\"at T\"ubingen, Germany\\
\email{werner@astro.uni-tuebingen.de}
\and
Institut f\"ur Physik der Atmosph\"are, DLR Oberpfaffenhofen, Germany\\
\email{Sabine.Miksa@dlr.de}
\and
Department of Physics and Astronomy, Johns Hopkins University,
Baltimore, MD 21218, U.S.A.\\
\email{kruk@pha.jhu.edu}
}
 
    \date{Received date; accepted date}
 
   \abstract{
We present a first systematic investigation of the iron abundance in very hot
(\Teff$\geq$50\,000\,K) hydrogen-deficient post-AGB stars. Our sample comprises
16 PG1159 stars and four DO white dwarfs. We use recent FUSE observations as
well as HST and IUE archival data to perform spectral analyses with line
blanketed NLTE model atmospheres. Iron is not detected in any PG1159 star. In
most cases this is compatible with a solar iron abundance due to limited
quality of HST and IUE data, although the tendency to an iron underabundance
may be recognized. However, the absence of iron lines in excellent FUSE spectra
suggests an underabundance by at least 1\,dex in two objects (\k1, NGC~7094). A
similar result has been reported recently in the [WC]-PG1159 transition
object Abell~78 (Werner \etal 2002). We discuss dust fractionation and
s-process neutron-captures as possible origins. We also announce the first
identification of sulfur in PG1159 stars.
\keywords{ 
           Stars: AGB and post-AGB --
           Stars: atmospheres --
           Stars: abundances -- 
           Stars: evolution --
           White dwarfs --
           Ultraviolet: stars
	 }}

\maketitle

\section{Introduction}

PG1159 stars are hot (\Teff=65\,000-180\,000\,K) hydrogen-deficient post-AGB
stars with carbon, helium and oxygen as main atmospheric constituents. They are
thought to be the result of a late helium shell flash (see e.g.\ Werner 2001).
Since they belong to the disk population and since diffusion effects can be
excluded due to ongoing mass-loss, one expects solar iron abundances.

Up to now no iron abundances were determined for these objects, for two
reasons. First, iron is highly ionized so that spectral lines have to be looked
for in the UV and FUV region. High resolution, high-S/N spectra are required
from these mostly faint objects. A few of the brightest were observed with IUE,
however, line identification is doubtful or impossible, as we will show. HST
spectra of many PG1159 stars are archived; however, in most cases the spectral
resolution is not high. In this paper we use all this archival material to make
a systematic investigation of iron lines with NLTE model atmospheres. We also
use new FUSE spectra of seven PG1159 stars and of the hottest known DO white
dwarf for this purpose. Second, spectral analyses require NLTE model
atmospheres. Models that include iron opacities for high ionization stages have
only recently become available.

\section{Sample selection and observational data}

Our sample comprises all PG1159 stars for which HST and IUE high resolution
archival spectra with a minimum acceptable S/N are available. These are 16 out
of the 35 objects known. FUSE spectra are available for seven of the program
stars. The sample is augmented by four hot non-DA (DO) white dwarfs in order to
complete an earlier analysis of DO spectra (Dreizler 1999). The twenty program
stars are listed in Table~\ref{objects_tab}. The spectra used for our analysis
are listed in Table~\ref{spectra_tab}.

\begin{table*}
\begin{center}
\caption{Atmospheric parameters of the program stars as taken from the
literature. The iron abundance is determined in this work, except for Abell~78.
Abundances are given in \% number fractions, except for iron. The iron
abundance in PG1159 stars is the logarithm of mass fraction relative to the
solar value. For the DO stars Fe is given as number ratio relative to
He. Pulsating PG1159 stars are marked with an asterisk\label{objects_tab}}
\begin{tabular}{l l r r r r r r r l l}
      \hline
      \noalign{\smallskip}
Object  & type      & \Teff & \logg & H & He & C & N & O & log Fe/Fe$_\odot$ & ref.\\
        &           & [kK]  &(cgs)  &   &    &   &   &   & (mass fraction)   &   \\
      \noalign{\smallskip}
      \hline
      \noalign{\smallskip}

\rxj           & PG1159$\ast$ & 170 &    6.0  &    &       61 &     31   &    &       8    &$\leq 0$&A\\
PG1520+525     & PG1159       & 150 &    7.5  &    &       72 &     21   &   $<0.01$ &  7  &$<-0.5$ &L\\
\ngc           & PG1159$\ast$ & 150 &    5.7  &    &       85 &     13   &  &  2           &$\leq 0$&H\\
PG1159-035     & PG1159$\ast$ & 140 &    7.0  &   $<5$   &   61 &     30  &    1    &   8  &$<-0.5$&K,L\\
\k1            & PG1159$\ast$ & 140 &    6.4      &   &    61   &   31    &  &        8    &$<-1.0$ &I\\   
HS2324+397     & PG1159$\ast$ & 130 &    6.2  &   61 &     30 &     9   &    $<0.1$ &   0.2&$<-0.5$ &D\\
Abell 43       & PG1159$\ast$ & 110 &    5.7  &   74 &     22 &     4   &    &             &$\leq 0$&G\\      
Abell 78       & PG1159       & 110 &    5.5  &    &       62 &     31  &    &        7    &$<-1.5$ &J,M\\
NGC\,7094      & PG1159$\ast$ & 110 &    5.7  &   74 &     22 &     4   &    &             &$<-1.5$ &G\\      
PG1424+535     & PG1159       & 110 &    7.0  &   &        76 &     22  &    $<0.001$&   2 &$\leq 0$&L\\
HS1517+7403    & PG1159       & 110 &    7.0  &   &        95 &     5   &    $<0.001$ & 0.5&$\leq 0$&L\\
PG2131+066     & PG1159$\ast$ & 95  &    7.5  &   &        71 &     21  &    1 &      7    &$\leq 0$&L\\
MCT0130-1937   & PG1159       &  90&     7.5  &   &        90 &     9   &    $<0.001$ &  1 &$\leq 0$&L\\
PG1707+427     & PG1159$\ast$ & 85 &     7.5  &   &        71 &     21  &    1  &     7    &$\leq 0$&L\\
PG0122+200     & PG1159$\ast$ & 80 &     7.5  &   &        71 &     21  &    1  &     7    &$\leq 0$&L\\
HS0704+6153    & PG1159       & 75 &     7.0  &   &        88 &     9   &    $<0.001$ &  3 &$\leq 0$&L\\
 \noalign{\smallskip}
      \hline
      \noalign{\smallskip}
               & &      &         &   &            &        &         &            & Fe/He& \\
               & &      &         &   &            &        &         &            & (number ratio)& \\
 \noalign{\smallskip}
      \hline
      \noalign{\smallskip}
\kpd          &DO&  120 &    7.0  &   &        100 &    0.1  & 0.01 &  $ <10^{-5}$      &$<10^{-5}$&A\\   
PG1034+001    &DO&  100 &    7.5  &   $<5$  &    100 &    0.001 &   0.063 & 0.0079      &$1\cdot10^{-5}$&B,C\\ 
HZ21          &DO&  53  &    7.8  &   $<10$ &    100 &    $<0.001$ &  $<0.001$&  $<0.001$&$<10^{-4}$&E\\
HD149499B     &DO&  50 &     8.0  &   18  &    82  &    0.001 &       &                 &$<10^{-5}$&E,F\\
 \noalign{\smallskip}
      \hline
     \end{tabular}
\end{center}
References in last column: 
A: Werner \etal 1996,
B: Werner 1996,
C: Werner et al. 1995,
D: Dreizler \etal 1996,
E: Dreizler \& Werner 1996,
F: Napiwotzki \etal 1995,
G: Dreizler \etal 1997,
H: Rauch \& Werner 1997,
I: Kruk \& Werner 1998,
J: Werner \& Koesterke 1992,
K: Werner \etal 1991,
L: Dreizler \& Heber 1998,
M: Werner et al. 2002
\end{table*}

\begin{table*}
\begin{center}
\caption{List of spectra for the program stars. The IUE spectra cover a
wavelength range of 1150-1980\,\AA\ with a resolution of about 0.1\,\AA. The
available FUSE spectra cover 905-1187\,\AA\ with a resolution of about 0.05\,\AA
\label{spectra_tab} }
\begin{tabular}{l| c c c c|  c c|  c c}
      \hline 
      \noalign{\smallskip}
Object &\multicolumn{4}{c}{| HST | }&\multicolumn{2}{c}{| IUE
      |}&\multicolumn{2}{c}{| FUSE |}\\
       & dataset & $\lambda$-range & res. & t$_{\mathrm{exp}}$ & SWP
        & t$_{\mathrm{exp}}$ & dataset & t$_{\mathrm{exp}}$ \\
       &    name      & [\,\AA\,]   & [\,\AA\,]     & [min] & number & [min] &name&[min]\\
 \noalign{\smallskip}
 \hline
 \noalign{\smallskip}
       \noalign{\smallskip}
       \rxj        & Z27J0206T &1227$-$1263& 0.06   & 13& 47556 & 360&P1320501 & 137\\
       & Z27J0207N &1273$-$1309& 0.06   & 11& 47563 & 362&&\\
       & Z27J0208T &1343$-$1379& 0.06   & 16& 55411 & 363 &&\\
       PG1520+525   & Z2T20104T &1165$-$1461& 0.60& 24&&& P1320101&81 \\
       \ngc           &&&&&  03353 & 112&& \\
       &&&&&  41997 & 120&& \\
       &&&&&  42068 & 223&& \\
       &&&&&  42073 & 115&& \\
       &&&&&  42104 & 150&& \\
       &&&&&  42214 & 150&& \\
       &&&&&  42247 & 120&& \\
       &&&&&  47843 & 165&& \\
       &&&&&  47844 & 155&& \\ 
       PG1159-035   &&&&&  23032 & 1020 &Q1090101&105\\
       &&&&&  53903 & 667 &&\\
       &&&&&  54675 & 875 &&\\
       &&&&&  54976 & 880 &&\\
       \k1          & Z1EJ0304M &1497$-$1532& 0.06& 13&&&I8110302&665\\
       & Z1EJ0305M &1525$-$1560&0.06& 13&&&&\\
       HS2324+397   & Z3GW0204T &1140$-$1436&0.60& 74&&&P1320601&67 \\
       Abell  43      & Z3GW0304T &1139$-$1435& 0.60   & 71 &  38955 &385& &\\
       Abell  78      &           &           &        &    &  16967 & 40& & \\
                      &           &           &        &    &  19879 & 425&&\\
                      &           &           &        &    &  19906 & 420&& \\
       NGC\,7094      & Z3GW0104T &1139$-$1435& 0.60   & 22& 52919 & 330&P1043701&386\\
       & Z3GW0105T &1455$-$1751& 0.60   & 47 & 56107 & 400 &&\\
  &&&&&  56112 & 400&& \\
      &&&& &  56120 & 400&& \\
       PG1424+535   & Z2T20204T &1166$-$1452& 0.60& 24 &&&&\\
       HS1517+7403  & Z2T20704T &1165$-$1451& 0.60& 24&&&&\\
       PG2131+066   & Z2T20804T &1165$-$1451& 0.60& 18 &&&&\\
       MCT0130-1937 & Z2T20504T &1165$-$1451& 0.60 & 17&&&& \\
       PG1707+427   & Z2T20304T &1165$-$1451&0.60& 20&&&P1320401 & 243\\ 
       PG0122+200   & Z2T20404T &1164$-$1450& 0.60& 18&&&& \\
       HS0704+6153  & Z2T20604T &1165$-$1461& 0.60& 27 &&&&\\
       & Z2T20605T &1165$-$1461& 0.60& 27 &&&&\\
\hline
       \kpd        & Z27J0106T &1227$-$1263& 0.06& 13 & 26191 & 420&M1070201&132\\
       & Z27J0107T &1273$-$1309&0.06& 11 & 52108 & 360             &M1070202&77\\
       & Z27J0108T &1343$-$1379& 0.06& 16 & 52146 & 425            &M1070203&138\\
       &           &           &     &    &  52185 & 425           &P1040101&114\\
       PG1034+001   & Z0YE0C08T &1185$-$1221& 0.06& 5 & 18509 & 330&&\\
       & Z0YE0C09T &1221$-$1256& 0.06& 10 & 26201 & 262&&\\
       & Z0YE0C0AT &1369$-$1406& 0.06& 5&&&&\\
       & Z0YE0C0CT &1532$-$1568& 0.06& 5&&&&\\
       & Z0YE0C0DT &1622$-$1658& 0.06& 5&&&&\\
       HZ21         & Z3GM0404T &1227$-$1265& 0.06& 42 & 31287 & 1147&& \\
       & Z3GM0405T &1339$-$1375& 0.06& 33 &&&&\\
       HD149499B     & Z3GM0504P &1227$-$1265& 0.06   & 7 & 06272 & 90&&\\
       & Z3GM0505P &1338$-$1375& 0.06   & 7 &  13781 & 40&& \\
       & Z3GM0506P &1699$-$1735& 0.06   & 9 & 13782 & 28&&\\
                      &           &           &        &     &  13783  & 89&& \\
                     &            &           &        &     &  17467  & 75 &&\\
      \hline
     \end{tabular}
\end{center}
\end{table*}

\subsection{FUSE}

The FUSE datasets in programs P132 and Q109 were observed for the present study
of PG1159 stars, the datasets in program P104 were observed for ISM studies,
and the datasets in program M107 were observed for the FUSE calibration program.
Each of these datasets is available from the Multimission Archive at the Space
Telescope Science Institute (MAST). All observations were obtained using the
30\arcsec$\times$30\arcsec\ LWRS spectrograph aperture. The data were obtained
in ``timetag'' mode, except for P1320501 and M1070201,2,3 datasets, which were
obtained in ``histogram'' (or spectral image) mode.  These datasets were all
reduced using the standard CALFUSE pipeline, version 1.8.7.  These observations
exhibit RMS variations in observed flux that are usually less than 1\% from one
exposure to another, indicating that no significant signal was lost as a result
of channel misalignments.  The one exception was SiC1 data for exposure 1 of
the observation of \rxj, for which the flux level was about 20\% below that of
the other exposures.  A description of the FUSE instrument and the channel
alignment issues are given by Moos \etal (2000) and Sahnow \etal (2000).  For
most objects the spectra obtained in different exposures were coaligned by
cross-correlating on regions containing narrow interstellar absorption features
before combining to produce a spectrum for an entire observation.  The
resulting spectral resolution was typically about 20\,\kms.  The spectra
obtained in single exposures for the objects HS2324+397 and PG1707+427 did not
have sufficient signal-to-noise ratios to permit cross-correlation, so these
spectra were combined without coalignment. In these cases the spectral
resolution was somewhat worse, about 25\,\kms\ for PG1707+427, and
25--30\,\kms\ for HS2324+397.

The observation of \k1\ required special processing.  This observation was
obtained during in-orbit checkout very early in the mission, as part of a test
to determine both the channel alignment offsets and the primary mirror to
spectrograph slit focus offset for each channel. The test consisted of slewing
the telescope in 40 1\arcsec\ steps during each exposure, centered on the LiF1
LWRS aperture.  There were a total of 11 scans each in the dispersion and
cross-dispersion directions.  Because the channels were only crudely aligned at
the time of this test, the effective exposure time varied for each channel,
from about 38~ksec for LiF1 to only 8~ksec for SiC2.  Data were discarded for
time periods in which the star was outside of the aperture, or for which the
count rate otherwise deviated significantly from the mean. These data were
processed using version 1.7.7 of CALFUSE, but the differences between this
version and 1.8.7 were not significant for the purposes of this program. Custom
processing was added to the pipeline to correct the recorded photon positions
for the slewing of the spacecraft across the spectrograph apertures during each
exposure.  The resulting resolution was about 24\,\kms.

\subsection{HST and IUE}

HST data was retrieved from the MAST. The spectra were taken with the Goddard
High Resolution Spectrograph (GHRS) with a resolution of 0.06\,\AA\ or
0.6\,\AA. All exposures were made in the FPSPLIT mode which results in four
individual spectra. As only the first exposure was wavelength-calibrated, the
subsequent exposures were adjusted in wavelength to match the first using
cross-correlation.

Reduced SWP spectra were extracted from the IUE Final Archive using preview
data. They cover a wavelength range of 1150--1980\,\AA, with a resolution of
about 0.1\,\AA. If more than one spectrum was available they were co-added
using the S/N-ratio as weighting factor.

\begin{figure}
\includegraphics[width=\columnwidth]{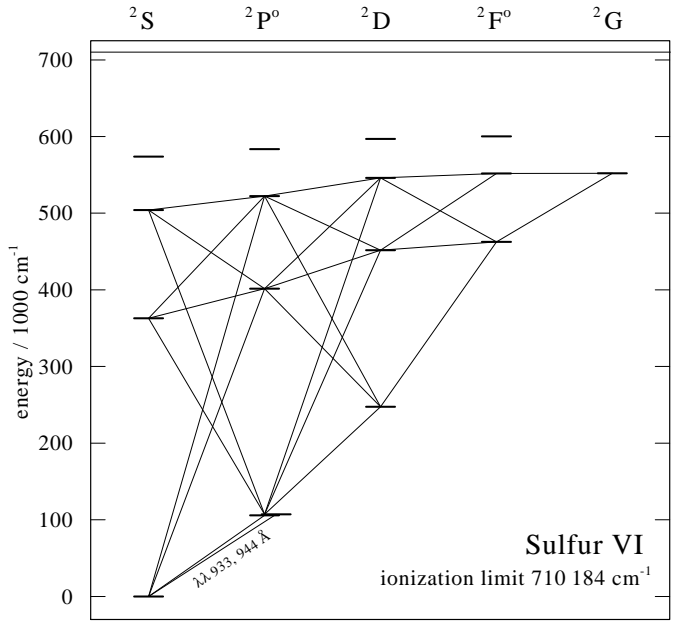}
\caption{ Model atom for \iion{S}{VI}. The transition responsible for the
\iion{S}{VI} resonance doublet in the FUSE range is labeled\label{grotrian} }
\end{figure}

\begin{figure}
\includegraphics[width=\columnwidth]{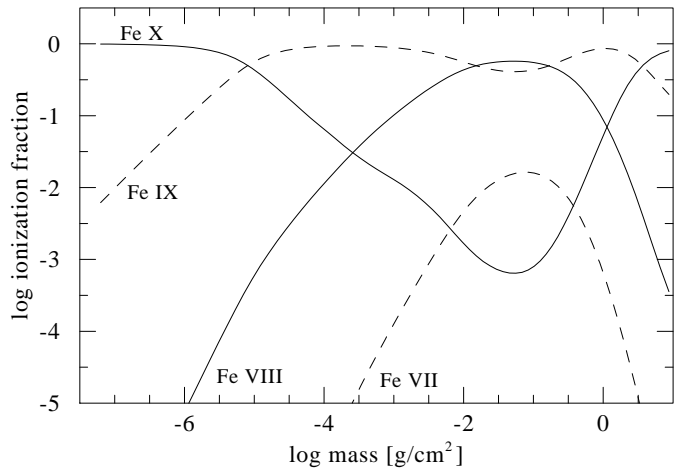}
\caption{Ionization structure of the model atmosphere for \k1\ with
\Teff\,=\,140\,000\,K, \logg\,=\,6.4 and a solar iron
abundance\label{K1-16_IONPLOT} }
\end{figure}

\begin{figure*}
\includegraphics[width=\textwidth]{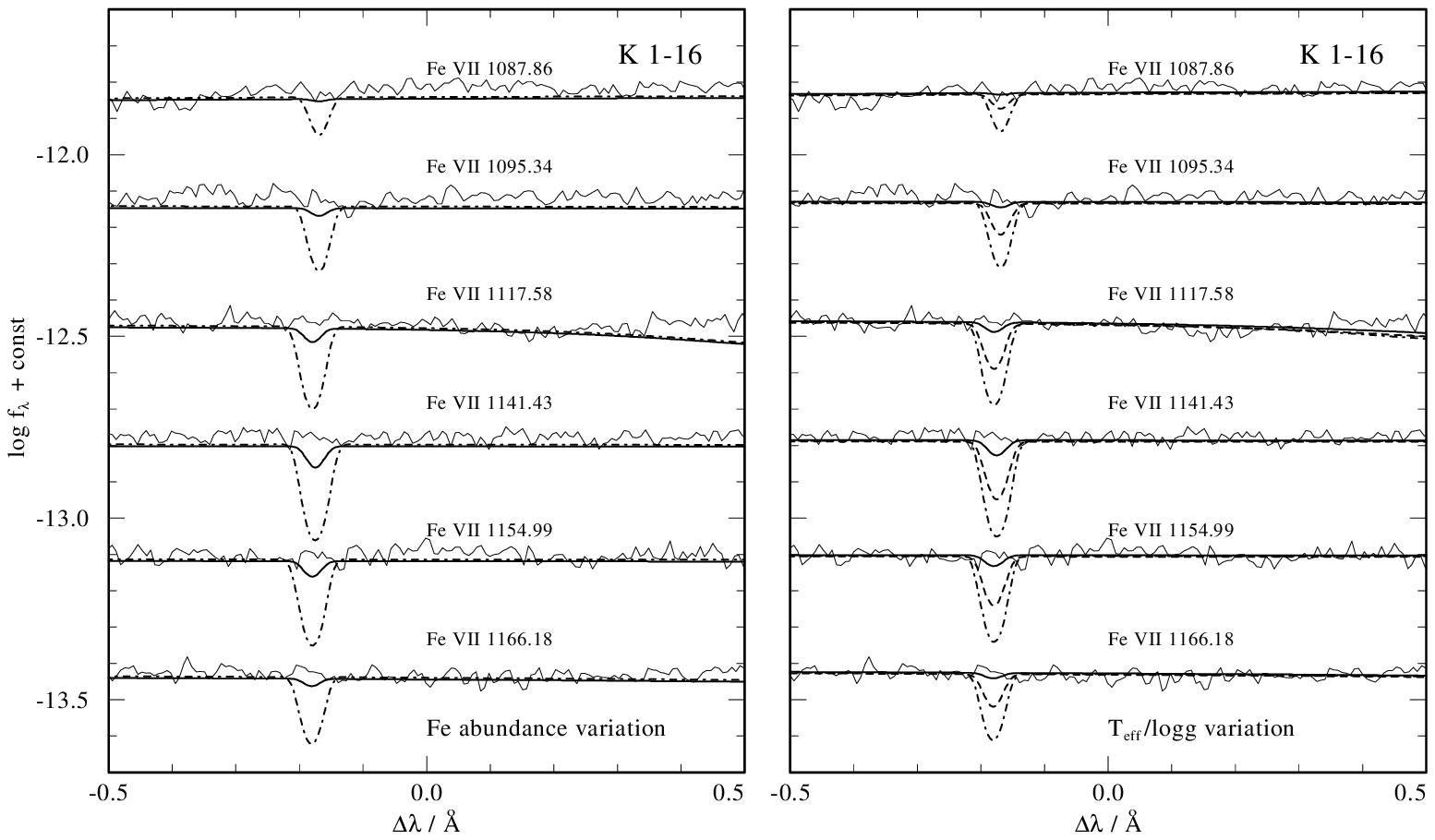}
\caption{The strongest iron lines expected in the FUSE data of \k1\ are from
\iion{Fe}{VII}. None of these can be identified in \k1\ (LiF2A data). Left
panel: Models with \Teff=140\,000\,K, \logg=6.4 and iron abundances of 0.1 solar
(solid) and solar (dash-dotted line). Right panel: Models with
\Teff=140\,000\,K/\logg=6.4 (dash-dotted), 150\,000/6.5 (dashed) and
160\,000/6.6 (solid), and solar iron abundance. A velocity shift of
$-$48.07\,\kms\ is taken into account according to Holberg \etal
(1998)\label{K1-16}. The spectra are smoothed with 0.01\,\AA\ Gaussians}
\end{figure*}

\section{Model atmospheres and synthetic spectra}

We have calculated a plane-parallel model atmosphere in radiative and
hydrostatic equilibrium for each program star with our ALI code (Werner \&
Dreizler 1999) which can handle line blanketing by the iron group elements. We
use the line list of Kurucz (1991). In general the models are similar to those
of Deetjen \etal (1999), so we will restrict ourselves to a summary of the
model atoms in Table~\ref{levels_tab}. However, higher ionization stages of
iron are taken into account due to the higher effective temperature of our
program stars. 

Sulfur line formation calculations were performed for the central star of \k1.
The sulfur model atom (\iion{S}{IV--VII}) has been constructed using the
Opacity Project (OP, Seaton \etal 1994) database which is a basic source for
energy levels, oscillator strengths, and photoionization cross-sections. Since
the OP energy levels are slightly different from laboratory measurements, we
have replaced these values by those listed in Bashkin \& Stoner (1975). The
Grotrian diagram in Fig.\,\ref{grotrian} shows the levels and line transitions
of the \iion{S}{VI} model ion.
 
\begin{table}
\caption{Summary of the model atoms used in the model atmosphere calculations.
The numbers in brackets give the individual line numbers summed into superlines
for the iron ions. Each model atom is closed by a single NLTE level
representing the highest ionization stage. These levels are not specified here
\label{levels_tab}}
\begin{tabular}{l l r r r}
      \hline
      \noalign{\smallskip}
element & ion & NLTE levels & lines&\\
  \noalign{\smallskip}
 \hline
      \noalign{\smallskip}
      H  & \mbox{\scriptsize I}   & 5  & 6 \\ 
      \noalign{\smallskip}
      He & \scriptsize I   & 1  & 0 \\
         & \mbox{\scriptsize II}  & 32 & 79\\ 
      \noalign{\smallskip}
      O  & \mbox{\scriptsize IV}  & 1  & 0\\
         & \mbox{\scriptsize V}   & 6  & 0\\
         & \mbox{\scriptsize VI}  & 36 & 102\\ 
         & \mbox{\scriptsize VII} & 1  & 0 \\
      \noalign{\smallskip}
      N & \mbox{\scriptsize IV}  & 6  & 4 \\
        & \mbox{\scriptsize V}   & 4  & 1\\
        & \mbox{\scriptsize VI} & 1  & 0 \\
      \noalign{\smallskip}
      C  & \mbox{\scriptsize III} & 1  & 0\\
         & \mbox{\scriptsize IV}  & 36 & 96 \\
         & \mbox{\scriptsize V}   & 1  & 0\\
      \noalign{\smallskip}
        Fe & \mbox{\scriptsize III}  & 7 & 25 & (301\,981) \\
            & \mbox{\scriptsize IV}   & 7 & 25 & (1\,027\,793)\\
            & \mbox{\scriptsize V}   & 7 & 25 & (793\,718)\\
            & \mbox{\scriptsize VI}  & 8 & 33 & (340\,132)\\
            & \mbox{\scriptsize VII} & 9 & 39 & (86\,504)\\
            & \mbox{\scriptsize VIII} & 7 & 27 & (8\,724)\\
            & \mbox{\scriptsize IX}   & 8 & 32 & (36\,843)\\
            \noalign{\smallskip}
 \hline
  \noalign{\smallskip}
total   & & 184&494&(2\,595\,695) \\
            \noalign{\smallskip}
 \hline
     \end{tabular}
\end{table}

\section{Iron abundance}

For most of the objects the quality of the spectroscopic data is not sufficient
to detect iron lines when compared with models with solar Fe abundance (all
abundance values are given in mass fractions unless otherwise noted). For these
objects we can only conclude that the iron abundance is less or equal solar
(Table~\ref{objects_tab}).

\subsection{PG1159 stars}

Before we discuss individual objects, we briefly go into those stars for which
a solar abundance of Fe cannot be strictly excluded.

For PG1707+427 the FUSE data is superior to the HST data, however, the S/N of
the FUSE spectrum is still too poor to detect potential \iion{Fe}{VI} and
\iion{Fe}{VII} lines. For PG1424+535, HS1517+7403, PG2131+066, MCT0130-1937,
PG0122+200, and HS0704+6153 only HST data exist, but their resolution (0.6\,\AA)
is too low. For Abell~43 IUE and HST spectra exist, but with insufficient S/N
and resolution, respectively. Hence, for all the objects mentioned in this
paragraph, Fe lines remain hidden in the spectra if the abundance is solar.

In contrast to our model predictions no \iion{Fe}{VI} or \iion{Fe}{VI/VII}
lines can be detected in the FUSE spectra of HS2324+397 and PG1520+525,
respectively. We derive a slight underabundance of 0.5 dex, but due to the
relatively low S/N we must accept that a solar Fe abundance cannot be excluded
rigorously. The HST data are not useful because of insufficient resolution
(0.6\,\AA).

For PG1159-035 four co-added IUE spectra were analyzed, with no clear detection
of potential \iion{Fe}{VII} lines. Also, \iion{Fe}{VII} lines are not detectable
in the FUSE spectra, hence, Fe is possibly underabundant by 0.5\,dex or more.

\subsubsection{\k1}

\begin{figure}
\includegraphics[width=\columnwidth]{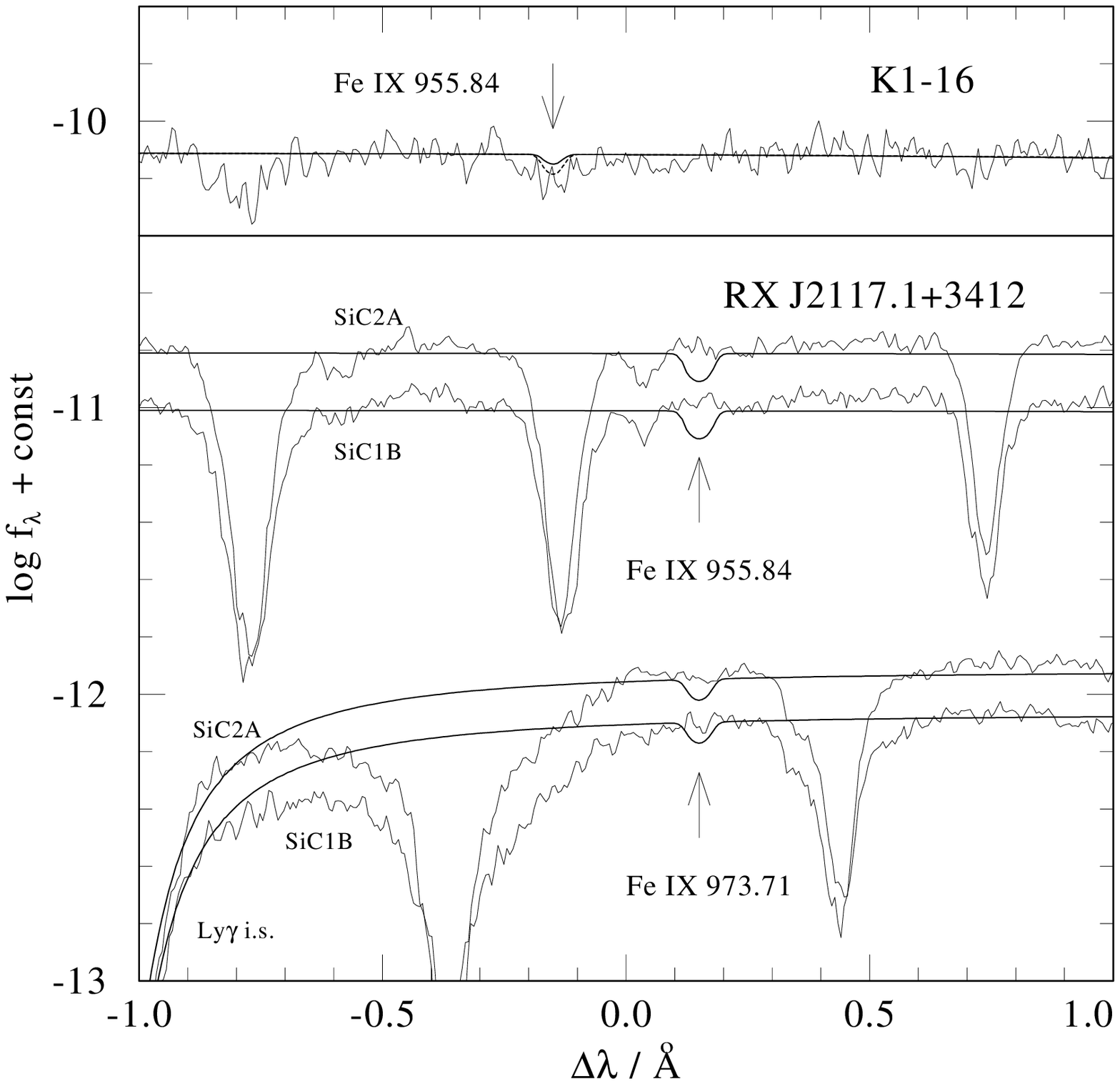}
\caption{Two \iion{Fe}{IX} lines are located in the FUSE range. Top panel: \k1\
compared to two model spectra (\Teff=140\,000\,K, \logg=6.4, full;
\Teff=160\,000\,K, \logg=6.6, dashed; solar Fe abundance) centered near the
\iion{Fe}{IX}~956\,\AA\ line. Bottom panel: \rxj\ compared to the model with
parameters as in Table~\ref{objects_tab} (\Teff=170\,000\,K, \logg=6.0, solar
Fe abundance) centered around \iion{Fe}{IX}~956\,\AA\ and \iion{Fe}{IX}~974\,\AA.
For this star both detector segments (SiC2A and SiC1B, slightly shifted
vertically) are shown. The spectra are smoothed with 0.01\,\AA\ Gaussians. A
radial velocity shift of +47.22\,\kms\ is taken into account\label{RXJ2117}}
\end{figure}

\k1\ is an extremely hot PG1159 type central star. Due to its high effective
temperature (140\,000\,K) the principal ionization stages of iron in the line
formation regions of the atmosphere are \iion{Fe}{VIII} and \iion{Fe}{IX}
(Fig.\,\ref{K1-16_IONPLOT}). A comparison of the excellent FUSE data with model
spectra reveals, that \iion{Fe}{VII} lines should be easily detectable if the
effective temperature is indeed 140\,000\,K (Rauch \& Werner 1997) and if the Fe
abundance is solar. Fig.\,\ref{K1-16}, however, shows that the strongest
\iion{Fe}{VII} model lines cannot be detected in the FUSE spectrum. Varying
\Teff\ and Fe abundance in the model suggests that either the Fe abundance is
lower than one tenth solar or that the effective temperature is higher than
160\,000\,K. The high \Teff\ would be surprising, because it means that the
quoted value, which was determined from optical spectra, would be in error by
at least 20\,000\,K. This is larger than a typical error of 10\% for such
analyses, however, the high \Teff\ cannot be excluded definitely. Hence we
conclude that an Fe underabundance of at least one dex is probably causing the
failure to detect \iion{Fe}{VII} lines in the FUSE data. Another argument for
the Fe underabundance can be given. Two \iion{Fe}{IX} lines with relatively
large gf-values are known, which are located in the FUSE spectral region,
contrary to most other lines of this ion, which are found in the 100--600\,\AA\
region. These lines (at 955.84\,\AA\ and 973.71\,\AA) clearly show up in the models
with solar Fe abundance, but nothing is seen in the FUSE spectrum at these
wavelengths (Fig.\,\ref{RXJ2117}). This holds particularly for the hotter model
(160\,000\,K), corroborating an underabundance even in the case that \Teff\ is
higher than previously found.

A final decision about the possibility of an Fe underabundance in \k1\ would
require a high-precision analysis of He, C, O line profiles to fix \Teff\ as
accurately as possible, using improved optical and UV spectroscopy in
combination with the FUSE data. Alternatively, the search for the bulk of
strongest \iion{Fe}{VIII} and \iion{Fe}{IX} lines in the 100-200\,\AA\ range could
be feasible with the Chandra observatory because \k1\ is a soft X-ray source
detected by ROSAT and EUVE.

The lack of \iion{Fe}{VII} lines in the \k1\ FUSE data supports a result by
Holberg \etal (1998). They found no evidence of \iion{Fe}{VII} lines in (much
less meaningful) IUE spectra, contrary to earlier work by Feibelman \&
Bruhweiler (1990).

\subsubsection{\rxj}

\rxj\ is the hottest known PG1159-type central star (\Teff=170\,000\,K). From
the discussion above it is clear, that the iron ionization is so high that
\iion{Fe}{VII} lines are not detectable, even in excellent data. This is indeed
true when inspecting the FUSE spectra. It is interesting to note that earlier
identifications of the \iion{Fe}{VII}~1239.69\,\AA\ line in a high resolution
HST/GHRS spectrum (Werner \etal 1996) and in IUE spectra (Feibelman 1999) are
therefore most likely wrong. Our model predicts that none of the potential
\iion{Fe}{VII} lines in the GHRS and IUE range appears. We believe that the
observed absorption is in fact an interstellar feature. There are two
\iion{Mg}{II} lines nearby, at 1239.93\,\AA\ and 1240.39\,\AA. The redshift of the
photospheric spectrum (+47.22\,\kms\ from Holberg \etal 1998, confirmed by our
measurement of the \iion{C}{IV}~1107.93\,\AA\ line) and the blueshift of the ISM
lines ($-$2.39\,\kms) conspire in a way that the interstellar \iion{Mg}{II}
1239.93\,\AA\ line almost coincides with the position of a possible stellar
\iion{Fe}{VII} line. The same problem arises in the case of \kpd\ (see below).
Generally, care must therefore be taken when iron is identified in hot stars by
means of this \iion{Fe}{VII}~1239.69\,\AA\ line alone. This can only be regarded
as safe when confusion with interstellar \iion{Mg}{II} can be excluded. The two
\iion{Fe}{IX} lines introduced above cannot be identified in the FUSE spectrum
(Fig.\,\ref{RXJ2117}). Like in the case of \k1\ this could hint at an Fe
underabundance and soft X-ray spectroscopy is required for a quantitative
analysis.

\begin{figure}
\includegraphics[width=\columnwidth]{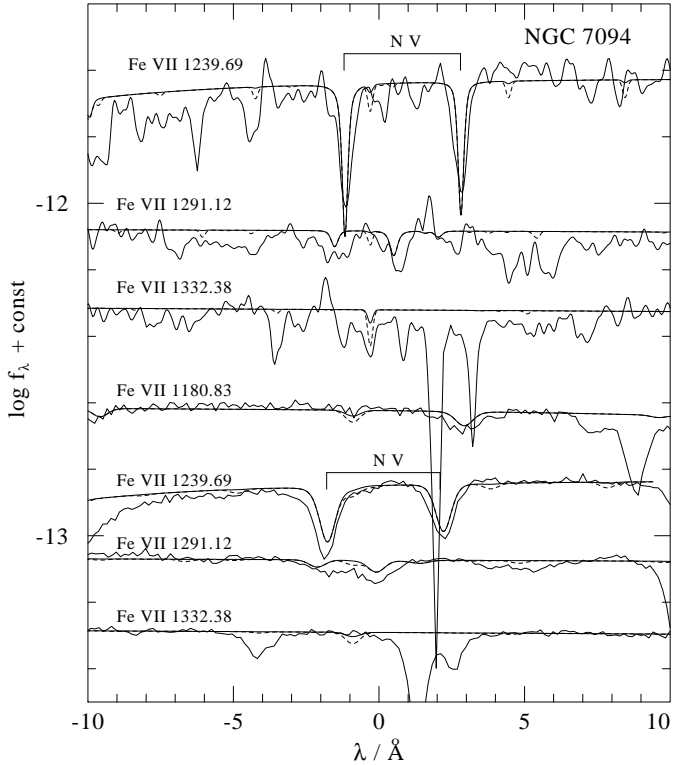}
\caption{The strongest \iion{Fe}{VII} lines in the spectra of NGC\,7094. The
top three lines are fitted to IUE spectra, the four lines at the bottom are
fitted to HST spectra. The models have iron abundances of 0.1 solar (solid
line) and solar (dashed line). No iron lines can be identified. The IUE and HST
spectra are smoothed 0.2\,\AA\ and 0.6\,\AA\ FWHM Gaussians, respectively. The
calculated spectra were shifted accounting for radial velocity. Note that the
radial velocity of the photospheric lines from HST data differs from that of
the IUE data \label{NGC7094}}
\end{figure}

\begin{figure}
\includegraphics[width=\columnwidth]{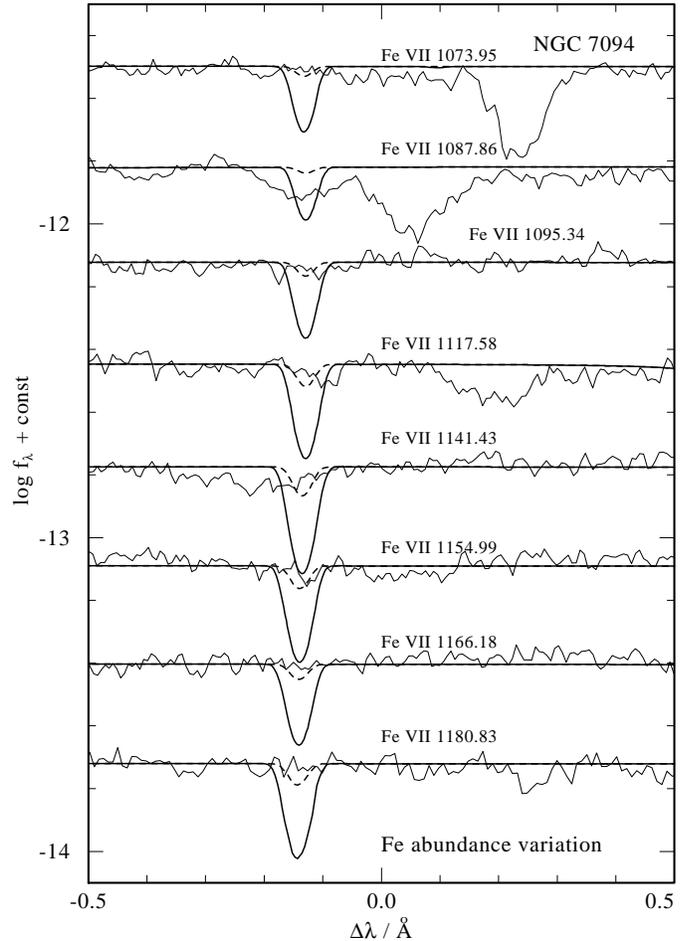} 
\caption{
Details of FUSE spectrum of NGC~7094 compared to theoretical line profiles,
centered around the strongest \iion{Fe}{VII} lines in the model. The model iron
abundances are 0.1 solar and 0.01 solar. No iron lines are detectable in the
FUSE spectrum. Spectra are smoothed with 0.01\,\AA\ Gaussians
\label{NGC7094_fuse}}
\end{figure}

\begin{figure}
\includegraphics[width=\columnwidth]{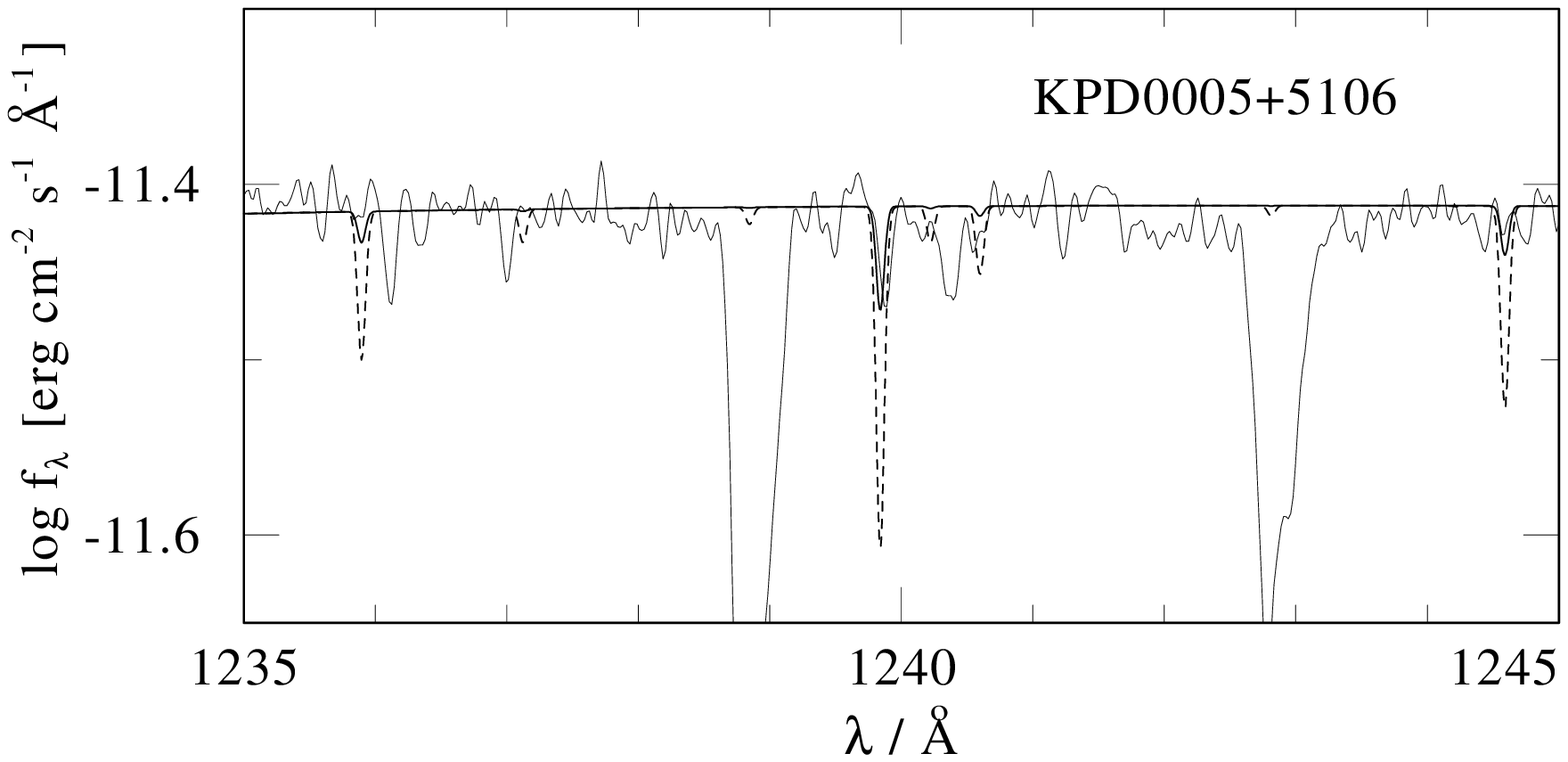}
\caption{The strongest \iion{Fe}{VII} lines expected in the HST spectrum of
\kpd. We show models with an iron abundance of 0.1 solar (solid line) and solar
(dashed line). No Fe lines can be identified. The feature at 1240\,\AA\ is
probably interstellar \iion{Mg}{II}. The HST spectrum is smoothed with 0.06\,\AA\
Gaussian. A radial velocity shift of +36.15\,\kms\ is taken into account
(Holberg \etal 1998) \label{KPD0005}}
\end{figure}

\subsubsection{\ngc}

\ngc\ is also a very hot (\Teff=150\,000\,K) PG1159 type central star. Numerous
IUE high resolution spectra exist which can be co-added. Holberg \etal (1998)
find no evidence of the \iion{Fe}{VII} lines reported by Feibelman \&
Bruhweiler (1990). We confirm this result and find that this is due to the high
temperature. Better spectra (from HST or FUSE) might be able to detect these
lines, which are weak but not completely absent in the solar Fe abundance
models. It appears that the tentative identification of lines from even lower
ionized \iion{Fe}{VI} by Feibelman \& Johansson (1995) and Feibelman (1995) is
very likely not correct.

\subsubsection{NGC\,7094}

NGC\,7094 is a so-called hybrid PG\,1159 star, meaning that detectable amounts
of hydrogen are present in its atmosphere. We fail to detect iron lines in the
IUE and HST spectra of NGC\,7094 (Fig.\,\ref{NGC7094}). This is in contrast to
Feibelman (2000), who claimed identification of several \iion{Fe}{VI} and
\iion{Fe}{VII} lines in the IUE data. This central star is so hot that our
model predicts stronger \iion{Fe}{VII} than \iion{Fe}{VI} lines. One can derive
an underabundance of 0.5\,dex. Inspecting the superior FUSE data we, again,
cannot identify \iion{Fe}{VII} lines. Comparison with our model reveals an even
higher underabundance of 1--2\,dex (Fig.\,\ref{NGC7094_fuse}).

\subsubsection{Abell 78}

The central star of the planetary nebula Abell~78 is one of the rare
[WC]--PG1159 transition object showing spectral signatures of both early [WC]
spectral type (emission lines) and PG1159 type (absorption lines). As in the
case of NGC\,7094 only the FUSE spectrum allows one to determine a strict limit
to the Fe abundance. A first analysis has been presented recently by Werner
\etal (2002) and the result is an underabundance of 1--2\,dex.

\subsection{DO white dwarfs}

\kpd\ is the hottest known DO white dwarf, showing peculiar emission lines due
to ultrahigh ionized light metals (Werner \& Heber 1992) and it is the only
white dwarf known to have an X-ray corona (Fleming \etal 1993). An upper limit
for the Fe abundance can be inferred from the GHRS spectrum (one tenth solar,
Fig.\,\ref{KPD0005}). As in the case of \rxj, the absorption feature near
1240\,\AA\ cannot stem from \iion{Fe}{VII}, because no other \iion{Fe}{VII}
lines, neither in the HST nor in the FUSE data, are detectable. We derive an
upper limit of log(Fe/He)$<-5$ (by number).

PG1034+001 is another hot DO white dwarf for which an iron abundance of
log(Fe/He)=$-$5 has been derived from \iion{Fe}{VI} lines in HST data (Werner
\etal 1995). We confirm this result by our re-analysis.

HZ~21 is one of the coolest DO white dwarfs. We confirm the result by Holberg
\etal (1998) that only a single photospheric line can be identified in the IUE
spectra (\iion{He}{II}~1640\,\AA). The HST spectra are superior in quality, but
they only cover wavelength ranges where \iion{Fe}{VI} and \iion{Fe}{VII} are
located. HZ~21 is too cool to ionize iron this strongly. We derive an upper
limit of log(Fe/He)$<-4$.

HD149499B is also a cool DO star. From the absence of \iion{Fe}{V} lines in the
HST spectrum we derive an upper limit for the Fe abundance: log(Fe/He)$<-5$.

\begin{figure}
\includegraphics[width=\columnwidth]{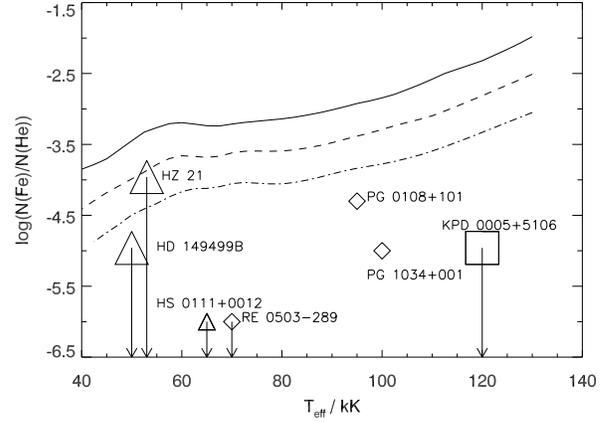}
\caption{Comparison of predicted iron abundances to our results for the DO
white dwarfs. The graphs show diffusion equilibrium abundances for iron at
$\tau=2/3$ as calculated by Chayer \etal (1995): \logg\,=\,7.0 (solid line),
\logg\,=\,7.5 (dashed), \logg\,=\,8.0 (dash-dotted). The symbols mark objects
with \logg\,=\,7.0 ($\Box$), \logg\,=\,7.5 ($\Diamond$) and \logg\,=\,8.0
($\triangle$). Small symbols mark the objects analyzed by Dreizler
(1999)\label{chayer_new}}
\end{figure}

To conclude our results on the DO stars, the iron abundance is much lower than
predicted by diffusion theory (Fig.\,\ref{chayer_new}), confirming earlier
results on other DO white dwarfs (Dreizler 1999).

\begin{figure*}
\includegraphics[width=\textwidth]{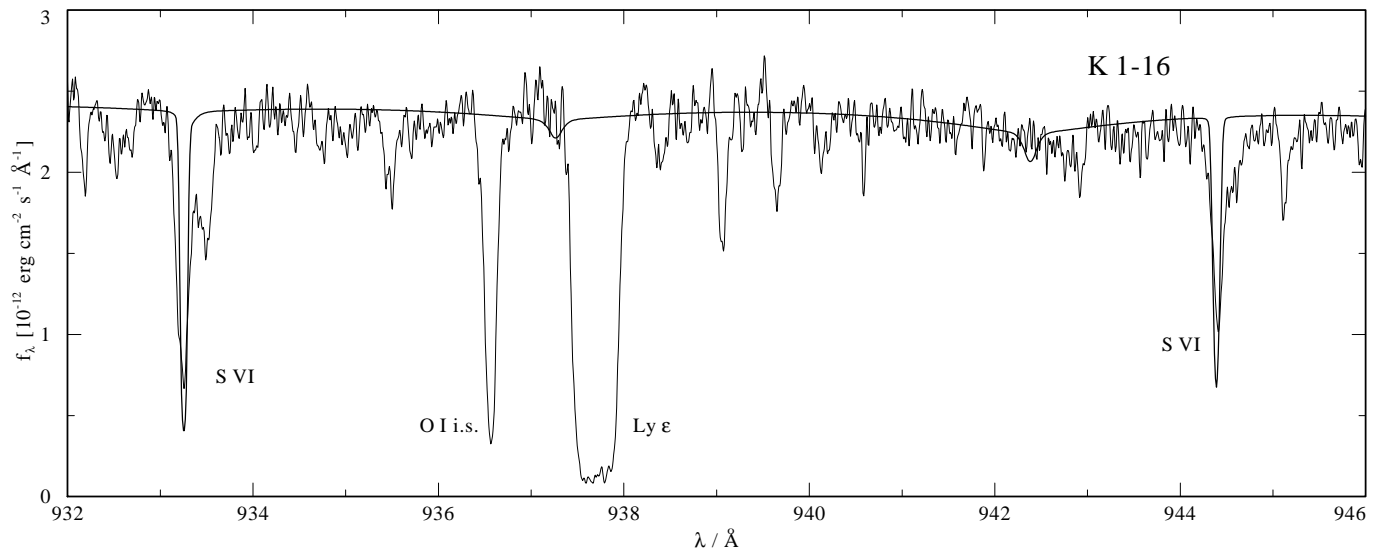} \caption{Line profile fit to
the \iion{S}{VI} resonance doublet in the FUSE spectrum of \k1. The model has a
solar sulfur abundance. Other parameters as in Table~\ref{objects_tab}. The
spectra are smoothed with 0.02\,\AA\ Gaussians\label{S_linien}}
\end{figure*}

\subsection{Summary of iron line analyses}

To summarize this section, we fail to identify iron lines in any PG1159 star.
In most objects this is compatible with a solar iron abundance. In three cases
(PG1159-035, HS2324+397, PG1520+525) we tentatively propose an interpretation
in terms of a slight iron underabundance (0.5\,dex). In three cases (\k1,
NGC\,7094, Abell~78) the underabundance is estimated to 1--2\,dex.

For three out of four DO white dwarfs investigated in this work we derive an
upper iron abundance limit from the absence of respective lines in their
spectra. In the case of PG1034+001 the analysis of detected iron lines
corroborates an earlier abundance determination.

\section{Detection of sulfur in \k1\ and other PG\,1159 stars}

The FUSE spectrum of \k1\ allows for the first identification of sulfur in a
PG1159 star based on the 933.38/944.52\,\AA\ resonance doublet
(Fig.\,\ref{S_linien}). It seems that both lines are split into two components,
especially the \iion{S}{VI} 933.38\,\AA\ line. The stronger blueshifted components
are attributed to the photosphere. Their velocity shift is about $-$40\,\kms\
which is in coarse agreement with measurements from IUE spectra ($-$48.07\,\kms,
Holberg \etal 1998). Our model fit suggests a solar sulfur abundance.

The \iion{S}{VI} doublet is also detected in the spectra of \rxj, PG1520+525,
and PG1159-035.

\section{Discussion}

Whether the iron deficiency in PG1159 stars is related to the same phenomenon
observed in metal-poor post-AGB B-A-F supergiants can only be speculated upon.
In some cases dust fractionation on the AGB has been invoked to explain the Fe
deficiency (Van Winckel \etal 1992), whereas in other cases evidence was found
that iron was transformed to heavier elements by s-process neutron captures
(Decin \etal 1998). To our opinion it is unlikely that dust fractionation on
the AGB is responsible for the iron deficiency in PG1159 stars, because strong
post-AGB mass-loss removes layers ``cleaned'' from iron during the AGB phase.

In the case of PG1159 stars the high C and O abundances result from envelope
mixing caused by a late He-shell flash (Herwig \etal 1999). This event also
modifies the near-solar abundance ratios of iron-peak elements in the envelope
by dredging up matter in which s-process elements were built-up by n-capture on
$^{56}$Fe seeds during the AGB phase. This scenario can be tested by analyzing
the resulting Fe/Ni abundance ratio, because it is significantly changed in the
intershell region in favor of Ni by the conversion of $^{56}$Fe into $^{60}$Ni.
The Fe depletion by n-captures typically amounts to a factor of 10 (Busso \etal
1999). In order to roughly estimate the Ni/Fe ratio one can assume nuclear
statistical equilibrium. The two most abundant Ni isotopes are $^{60}$Ni (26\%)
and $^{58}$Ni (68\%). During s-process $^{58}$Ni is destroyed (and not
synthesized), by conversion into $^{60}$Ni. $^{60}$Ni is converted to heavier
elements 4 times faster than it is produced from $^{56}$Fe. Consequently, a
ratio Fe/Ni$\approx$4 results, which is a factor of five below the solar value.
This could be detected by high resolution HST and FUSE UV spectroscopy of
PG1159 stars, but only in objects that are cooler than the three hot stars for
which the strong Fe deficiency has been found. This is because in the hot
PG1159 stars \iion{Ni}{VII} lines are the prevailing nickel features but they
are located in the inaccessible EUV region. Interestingly, Asplund \etal (1999)
have indeed found that in Sakurai's object, which is thought to undergo a late
He-shell flash, Fe is reduced to 0.1 solar and Fe/Ni$\approx$3. This and other
s-process signatures might also be exhibited by Wolf-Rayet central stars and
PG1159 stars.

More quantitative results from nucleosynthesis calculations in appropriate
stellar models have been presented recently (Herwig \etal 2002) and inclusion of
nuclear networks in evolutionary model sequences will become available in the
near future.

Most recent results presented at the IAU Symposium 209 (Planetary Nebulae,
Canberra) confirm that iron deficiency among H-deficient post-AGB stars is not
restricted to PG1159 stars, as can be expected from evolutionary
considerations. As already mentioned, the [WC]-PG1159 transition object
Abell~78 is iron deficient (Werner \etal 2002) and three Wolf-Rayet central
stars are iron deficient, too. Gr\"afener \etal (2002) report a low Fe
abundance in SMP\,61, an early type [WC5] central star in the LMC. Its
abundance is at least 0.7\,dex below the LMC metallicity. Crowther \etal (2002)
find evidence for an iron underabundance of 0.3--0.7\,dex in the Galactic [WC]
stars NGC~40 ([WC8]) and BD+30\,3639 ([WC9]).

A general iron deficiency among PG1159 stars would also have implications for
asteroseismology. Roughly half of all PG1159 stars are GW\,Vir pulsators and
frequency analyses have revealed interesting results about the interior
structure of these stars (e.g.\ Winget \etal 1991). There is still a debate
about details of the pulsation driving mechanism. While it is accepted that
cyclic ionization of carbon and oxygen just beneath the photosphere is the main
driver, the iron opacity also does play a role (Saio 1996). In light of our
iron abundance analysis of the pulsator \k1\ it would be interesting to
re-address the problem of pulsation driving.

\begin{acknowledgements}
We thank Falk Herwig for useful discussions. HST data analysis in T\"ubingen is
supported by the DLR under grant 50\,OR\,9705\,5. JLD is supported by the DFG
under grant We 1312/23-1.
\end{acknowledgements}

\end{document}